\documentstyle[aps,epsf,epsfig,prl,multicol]{revtex}
\begin{document}
\draft
\title{Quantum Arnol'd Diffusion in a Simple Nonlinear System}
 \author{V.Ya. Demikhovskii$^1$, F.M. Izrailev$^2$ and A.I. Malyshev$^1$}
 \address{$^1$ Nizhny Novgorod State University, Gagarin Av., 23, 603950,
Nizhny Novgorod, Russia}
 \address{$^2$ Instituto de F\'isica, Universidad Aut\'onoma de Puebla,
 Apartado Postal J-48, 72570, Puebla, Mexico}
 \maketitle
\begin{abstract}
We study the fingerprint of the Arnol'd diffusion in a quantum
system of two coupled nonlinear oscillators with a two-frequency
external force. In the classical description, this peculiar
diffusion is due to the onset of a weak chaos in a narrow
stochastic layer near the separatrix of the coupling resonance. We
have found that global dependence of the quantum diffusion
coefficient on model parameters mimics, to some extent, the
classical data. However, the quantum diffusion happens to be
slower that the classical one. Another result is the dynamical
localization that leads to a saturation of the diffusion after
some characteristic time. We show that this effect has the same
nature as for the studied earlier dynamical localization in the
presence of global chaos. The quantum Arnol'd diffusion represents
a new type of quantum dynamics and can be observed, for example,
in 2D semiconductor structures (quantum billiards) perturbed by
time-periodic external fields.

\vspace{0.2cm}

\end{abstract}

\begin{multicols}{2}
\narrowtext

\section{Introduction}
As is well known, dynamical chaos in Hamiltonian systems is
related to the destruction of separatrices of nonlinear
resonances. For a strong interaction between the resonances, the
latter can overlap in the phase space, thus leading to a {\it
global chaos} for which chaotic region is spanned over the whole
phase space of a system, although large isolated islands of
stability may persist. For a weak interaction, chaotic motion
occurs only in the vicinity of separatrices of the resonances, in
accordance with the Kolmogorov-Arnol'd-Moser (KAM) theory (see,
for example, \cite{LL92}).

In the case of two degrees of freedom ($N=2$), a passage of the
trajectory from one stochastic region in phase space to another is
blocked by KAM surfaces. The situation changes drastically in
many-dimensional systems ($N>2$), where the KAM surfaces no longer
separate one stochastic region from another, and {\it chaotic
layers} of destroyed separatrices form a {\it stochastic web} that
can cover the whole phase space. Thus, if trajectory starts inside
the stochastic web, it can diffuse throughout the phase space.
This weak diffusion along stochastic webs was predicted by Arnol'd
in 1964 \cite{A64}, and since that time it is known as a very
peculiar phenomenon, universal for nonlinear Hamiltonian systems
with $N>2$.

There are many physical systems which behavior appears to be
strongly affected by the Arnol'd diffusion. As examples, one
should mention three-boby gravitational systems \cite{sanpaolo}
and galaxies dynamics \cite{gal}. It is argued that the Arnol'd
diffusion may have strong impact for the behaviour of our solar
system, it is also responsible for a loss of electrons in magnetic
traps (see discussion and references in \cite{C79}). From the
practical point of view, the Arnol'd diffusion may be dangerous
for long-time stability of motion of charge particles in high
energy storage rings
\cite{particle}.

Recently, the Arnol'd diffusion was explored in the classical
description of Rydberg atoms placed in crossed static electric and
magnetic fields \cite{MDU96}. The semiclassical approach has been
used for the stochastic pump model \cite{LW97}, where the effect
of quantization of the Arnol'd diffusion in a system of two pairs
of weakly coupled oscillators has been investigated.

For the first time, the Arnol'd diffusion was observed in
numerical study of a $4$-dimensional nonlinear map \cite{CKS71}.
The more physical model of two coupled nonlinear oscillators with
time-dependent perturbation was considered, both analytically and
numerically, in \cite{GIC77,old} (see also review
\cite{C79} and the books \cite{LL92}). Numerical experiments
with this model have confirmed analytical estimates obtained for
the diffusion coefficient in dependence on model parameters (for
recent studies on this subject see \cite{vita}). Note that direct
numerical study of the Arnol'd diffusion is quite difficult since
its rate is exponentially small, and it occurs only for initial
conditions inside vary narrow stochastic layers.

One should distinguish stochastic web for the Arnol'd diffusion
from that found for systems which are linear in the absence of a
perturbation, see for example \cite {F01}. In the latter case the
stochastic web arises around a {\it single} nonlinear resonance
that has infinite number of cells in the phase space (see
\cite{zasl} and references therein).

So far, the studies of the Arnol'd diffusion are restricted by the
classical (or semiclassical) approaches. On the other hand, it is
important to understand the influence of quantum effects. The
problem is not trivial since strong quantum effects can completely
suppress weak diffusion along narrow stochastic layers, even in a
deep semiclassical region \cite{S76}. The purpose of this work is
to study the fingerprint of the Arnol'd diffusion in a quantum
model, by making use a direct numerical simulation of some
nonlinear system, both in classical and quantum description.
Preliminary data are reported in Ref.\cite{we}.

The paper is organized as follows. In Sec. II the basic model is
introduced and discussed. We describe shortly the mechanism of the
classical Arnol'd diffusion in classical system and give the
expression for the diffusion coefficient. Sec. III is devoted to
the study of the quantum model. First, in Sec. III(A) we study the
structure of eigenstates of the stationary model (without an
external field). Second, in Sec. III(B) we show how to construct
the evolution operator that allows us to investigate the dynamics
of the system. Here we also discuss global properties of
quasienergy states. Next step of our consideration (Sec. III(C))
is the study of the evolution of the system for different initial
conditions and model parameters. We show that for initial states
corresponding to stochastic layer near the separatrix of the
coupling resonance, the motion has a diffusion-like character. We
calculate quantum diffusion coefficient and compare it with the
classical one. Quantum effects of the dynamical localization and
suppression of the classical diffusion are discussed in Sec.
III(D). In Sec.IV we give our conclusions by summarizing main
results, and shortly discuss possible systems in which the quantum
Arnol'd diffusion may be observed.

\section{Classical model}
In this Section we discuss main results on the Arnol'd diffusion
obtained in \cite{GIC77} for a classical model. The Hamiltonian of
this model describes two nonlinear oscillators coupled by the
linear term, and governed by an external force $f(t)$,
\begin{equation}
\label{cl_ham_xy}
H=\frac{p_x^2}2+\frac{p_y^2}2+\frac{x^4}4+\frac{y^4}4 -\mu
xy-xf(t).
\end{equation}
Here $p_x$ and $p_y$ are momentums in the $x$ and $y$ directions,
and $\mu$ is the coupling constant. The driving term consists of
two harmonics of the same amplitude $f_0$,
\begin{equation}
\label{force}
f(t)=f_0(\cos{\Omega_1t}+\cos{\Omega_2t}),
\end{equation}
with commensurate frequencies,  $m\Omega_1=n\Omega_2$, so that the
period is $T=2\pi n/\Omega_1=2\pi m/\Omega_2$.

Without the coupling ($\mu=0$) and in the absence of the
perturbation ($f_0=0$), the motion of each oscillator is
integrable and can be found analytically. The quartic form of the
potentials has been chosen in order to have simple analytical
expressions in comparison with more realistic models with
additional quadratic terms $x^2/2$ and $y^2/2$ in the Hamiltonian.

An interesting feature of the system of quartic oscillators is a
small contribution of higher harmonics in spite of a strong
nonlinearity. Indeed, the solution for $x(t)$ has the form (for
details see \cite{C79}), $${x(t)\over a}={\rm cn}\,(\omega
t)={\pi\sqrt{2}\over K(1/\sqrt{2})}
\sum^{\infty}_{n=1}{\cos{\Bigl[(2n-1)\omega t\Bigr]} \over
\cosh{\Bigl[\pi(n-1/2)\Bigl]}}\approx $$
\begin{equation}
\label{resh}
0.9550\cos{\omega t} + {\cos{3\omega t}\over 23} + {\cos{5\omega
t}\over 23^2} + \dots.
\end{equation}
where $a$ is the amplitude of oscillations and $K(1/\sqrt 2)$
stands for the complete elliptic integral of the first kind. One
can see that the amplitude $a_m$ of higher harmonics sharply
decreases with an increase of $m$. Therefore, in action-angle
variables $I_x,\Theta_x$ one can approximately represent the
position of the $x-$oscillator by the expression $x \approx
a(I_x)\cos{\Theta_x}$. Then the system under consideration is
described by the Hamiltonian
\begin{equation}
\label{cl_ham}
\matrix{
H=A(I_x^{4/3}+I_y^{4/3})- \cr -\mu
a(I_x)a(I_y)\cos{\Theta_x}\cos{\Theta_y}-a(I_x)\cos{\Theta_x}f(t),}
\end{equation}
with $A=\Bigl( {3\pi\over4\sqrt{2}K(1/\sqrt{2})}\Bigr)^{4/3}$ and
$a(I_{x,y})= (4A)^{1/4}I_{x,y}^{1/3}$.

Near the {\it coupling resonance} $\omega_x=\omega_y$ the
resonance phase $\Theta_x-\Theta_y$ and amplitudes $a_x$, $a_y$
oscillate. Thus, it is convenient to introduce slow
($\theta_1=\Theta_x-\Theta_y$) and fast
($\theta_2=\Theta_x+\Theta_y$) phases by making use of the
canonical transformation with the generating function,
\begin{equation}
\label{gen}
F=(\Theta_x-\Theta_y)I_1+(\Theta_x+\Theta_y)I_2.
\end{equation}
As a result, new actions $I_1$ and $I_2$ are expressed as follows,
\begin{equation}
\label{action}
I_1={I_x-I_y\over 2}, \qquad I_2={I_x+I_y\over 2}.
\end{equation}
For the coupling resonance we have $I_x\approx I_y$, hence,
$I_1\ll I_2$, and the resonance Hamiltomian $H_{res}$ gets the
form,
\begin{equation}
\label{res_ham}
\matrix{
H_{res}=2AI_2^{4/3}+{B\over
2}I_1^2-V(\cos{\theta_1}+\cos{\theta_2})-
\cr -f_0a(I_2)\cos{\bigl( {\theta_1+\theta_2\over 2}\bigr)}
(\cos{\Omega_1t}+\cos{\Omega_2t}),}
\end{equation}
with $B(I_2)=8/9\cdot AI_2^{-2/3}$ and $V(I_2)=\mu a^2(I_2)/2$.

The average of Eq.(\ref{res_ham}) over the fast phase $\theta_2$
gives the \lq\lq pendulum\rq\rq{} Hamiltonian,
\begin{equation}
\label{res_ham_st}
\tilde H=2AI_2^{4/3}+{B\over 2}I_1^2-V\cos{\theta_1}.
\end{equation}
It defines the frequency $\tilde\omega$ of small oscillations of
$\theta_1$ and $I_1$,
\begin{equation}
\label{fr}
\tilde\omega=\sqrt{BV}=\beta \sqrt{\mu}, \,\,\,\,\,\,\,
\beta={\pi\over 2K(1/\sqrt{2})}\approx 0.85,
\end{equation}
and the half-width $\Delta\omega$ of the coupling resonance,
\begin{equation}
\label{hwc}
\Delta\omega=\beta \sqrt{2\mu}.
\end{equation}
Comparing Eqs.(\ref{res_ham}) and (\ref{res_ham_st}), one can
understand that the time-dependent perturbation destroys the
separatrix of the coupling resonance, and gives rise to a chaotic
motion in the vicinity of the separatrix. Specifically, the action
$I_2$ reveals a weak Arnol'd diffusion {\it along} the coupling
resonance {\it inside} the stochastic layer.

Thus, the long-term dynamics we are interested in, is controlled
by three resonances, the coupling and two driving ones. The
first-order driving resonances are determined by the condition
$\omega_x(I_x)=\Omega_1$, $\Omega_2$ where $\omega_x=4A/3\cdot
I_x^{1/3}$. In fact, the coupling resonance $\omega_x=\omega_y$
serves as a {\it guiding resonance} along which the diffusion
takes place.

All three resonances are characterized by their widths, they can
overlap with each other if the coupling constant $\mu$ and
perturbation strength $f_0$ are large enough. In order to observe
the Arnol'd diffusion, one needs to avoid such an overlap since it
leads to a strong {\it global chaos}. The condition for the
overlap of the resonances reads
\begin{equation}
\label{over}
\Delta\omega_1+{\Delta\omega\over\sqrt{2}}\cdot 2
+\Delta\omega_2\ge\delta\Omega,
\end{equation}
where
\begin{equation}
\label{hwd}
\Delta\omega_i=\beta \sqrt{2f_0\over a}
\end{equation}
is the half-width of the $i$-th driving resonance,
$\delta\Omega=|\Omega_1-\Omega_2|$. From Eqs.(\ref{hwc}) and
(\ref{hwd}) one can obtain for the overlap,
\begin{equation}
\label{over-2}
\sqrt{2f_0\over a}+\sqrt{\mu}\ge{\delta\Omega\over 2\beta}.
\end{equation}
The Arnol'd diffusion occurs in the case when inequality
(\ref{over-2}) does not satisfy.

In principal, the Arnol'd diffusion arises in our model even for
one driving resonance. However, in this case, the rate of the
diffusion will be strongly dependent on the distance between the
position of a trajectory inside the stochastic layer, and the
driving resonance in the frequency space. Instead, for two driving
resonances the Arnol'd diffusion is almost homogeneous if one
starts in between the two driving resonances. This simplifies the
analytical treatment of the diffusion, that has been performed in
Refs.\cite{GIC77}. Leaving aside technical details, we briefly
comment below the approach used in \cite{GIC77}.

In order to obtain the diffusion coefficient for a diffusion along
the coupling resonance, one needs to find the change of the total
Hamiltonian over the half-period of the unperturbed motion near
separatrix. Therefore, the diffusion coefficient can be evaluated
as follows,
\begin{equation}
\label{D}
D={\overline{(\Delta H)^2}\over T_a},
\end{equation}
where $T_a$ is the averaged period of motion within the separatrix
layer. The change of the Hamiltonian depends on an initial phase,
however, successive values of the phase can be treated as random
and independent. The variation of the total energy is then
determined by the sum over many periods for which successive
phases can be obtained via the {\it separatrix map}. Analytical
estimates obtained in \cite{GIC77}, give the following expression
for the diffusion coefficient (in action),
\begin{equation}
\label{Dact}
{\rm D}_I={af_0\over T_a \tilde\omega}\cdot{w_s^2\over\lambda^4}.
\end{equation}
Here $w_s=4\pi\nu\lambda^2 e^{-\pi\lambda/2}$ is the half-width of
chaotic layer of the coupling resonance, that mainly depends on
the adiabaticity parameter $\lambda= \delta\Omega/2\tilde\omega$,
with $\tilde\omega$ determined by Eq.(\ref{fr}).
\begin{figure}[tb]
\begin{center}
\mbox{\psfig{file=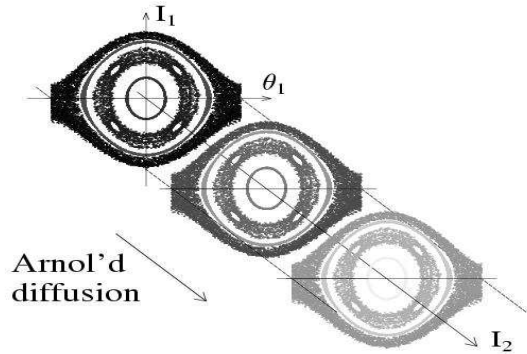,width=8cm,height=6cm}}
\end{center}
\caption{Three sections ($I_1, \theta_1$) of the phase space
for different values of $I_2$ are shown for the Hamiltonian
(\ref{cl_ham}) with the force (\ref{force}) with $\mu=2\cdot 10^{-4}$,
$f_0=2\cdot 10^{-6}$. The Arnol'd diffusion takes place along the
$I_2$-direction, see details in the text.}
\label{diagram}
\end{figure}

We performed numerical study of the classical Arnol'd diffusion in
the system described by the Hamiltonian (\ref{cl_ham}). The
frequencies of the perturbation (\ref{force}) were chosen as
$\Omega_1=0.2094$ and $\Omega_2=0.2513$ resulting in the period
$T=150$. Therefore, $\omega=(\Omega_1+\Omega_2)/2=0.23035$ which
determines the amplitude $a\approx 0.2719$. Correspondingly, the
initial conditions were taken for the system to be in between the
two driving resonances.

To put the system inside the stochastic layer of the coupling
resonance it is necessary to take $\theta_1=\pm\pi$. As in
Ref.\cite{GIC77}, the ratio $f_0/\mu=0.01$ was taken small enough
in order to avoid the overlapping of three first-order resonances
(see Eq.(\ref{over-2})). This also suppresses the influence of
second-order resonances between unperturbed non-linear motion and
the external perturbation.

Schematic structure of the coupling resonance is shown in Fig.1.
Numerical data are obtained for different initial conditions
corresponding to the separatrix layer and to the resonance region
of the coupling resonance. The chaotic region inside the resonance
is due to second-order resonances between non-linear motion of the
unperturbed Hamiltonian and two-frequency perturbation. The
condition of secondary resonances is
$n\cdot\tilde{\omega}'=m\cdot\delta\Omega$. Here $\tilde{\omega}'$
is the frequency of oscillations at the coupling resonance (near
the resonance centre we have
$\tilde{\omega}'\to\tilde{\omega}=\beta\sqrt{\mu}$), and $n,m$ are
integers. One can find that the stochastic region inside the
resonance corresponds to $m=1$ and $n=4$. As one can see, weak
diffusion along the coupling resonance can occur both within the
separatrix layer and inside the chaotic region formed by
second-order resonances.

The diffusion coefficient was computed as follows,
\begin{equation}
\label{Dform}
{\rm D}_n=\overline{(\Delta \bar H)^2 \over 10^n \cdot T}\, .
\end{equation}
Here $\bar H$ is the value of the total Hamiltonian
(\ref{cl_ham}), averaged over time intervals $T_n$ of length $10^n
\cdot T$ with $n=2,3$. The second average in (\ref{Dform}) has
been done in the following way. Having the mean value $\bar H$ in
each interval $T_n$ for a fixed $n$, we computed the difference
$\Delta
\bar H$ between adjacent intervals, and averaged the variance
$(\Delta \bar H)^2$ over all differences. This procedure was taken
in order to suppress large fluctuations of the energy, and to
reveal a stochastic character of motion. Specifically, in the case
of a true diffusion, one should expect $ \rm D_2\approx \rm D_3$.

\begin{figure}[tb]
\begin{center}
\mbox{\psfig{file=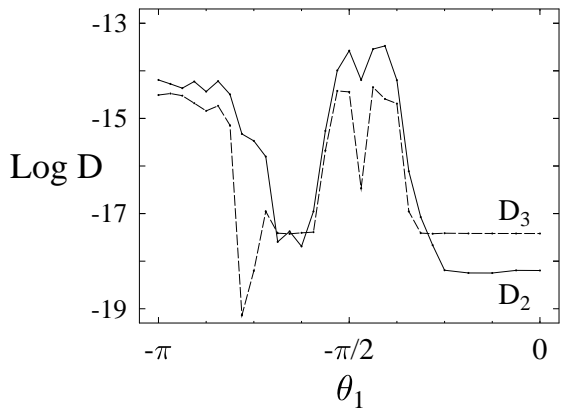,width=8cm,height=6cm}}
\end{center}
\caption{Dependence of the diffusion coefficient on initial
conditions inside the coupling resonance, see details in the text.
}
\label{cl_th_2}
\end{figure}

The dependence of diffusion coefficients $D_2$ and $D_3$ on
initial phase $\theta_1$ for $I_1=\theta_2=0$ and
$I_2=\omega^3/(3\beta^4)$ at $t=0$ is shown in Fig.\ref{cl_th_2}.
The region close to $\theta_1=-\pi$ corresponds to initial
conditions inside the separatrix layer, and the interval near
$\theta_1=-\pi/2$ corresponds to initial conditions inside the
inner stochastic region (see Fig.\ref{diagram}). In both these
regions Arnol'd diffusion coefficients have the same order.
Approximate equality $\rm D_2\approx \rm D_3$ indicates here that
the motion is really diffusion-like \cite{GIC77}. On the other
hand, strong difference between ${\rm D}_2$ and ${\rm D}_3$ in the
region $|\theta_1|<\pi/4$ manifests that the dynamics of the
system is non-diffusive.

% As was shown by Vecheslavov [\cite{Vech}] the effect of multifrequency
% perturbation, namely, low-frequency harmonics of this perturbation can
% define also the width of the main stochastic layer.

\section{Quantum model}

The corresponding quantum model is described by the Hamiltonian
(compare with Eq.(\ref{cl_ham_xy})),
\begin{equation}
\label{ham}
\hat H=\hat H^0_x+\hat H^0_y-\mu xy-f_0 x(\cos\Omega_1 t+\cos\Omega_2
t).
\end{equation}
Here
\begin{equation}
\label{hatH^0_x}
\hat H^0_x=\frac{\hat p_x^2}2+\frac{x^4}4,
\qquad \hat H^0_y=\frac{\hat p_y^2}2+\frac{y^4}4,
\end{equation}
and standard relations for momentum and coordinate operators are
assumed,
\begin{equation}
\label{commutator}
[\hat p_x,\,x]=-i\hbar_0, \qquad [\hat p_y,\,y]=-i\hbar_0
\end{equation}
with the dimensionless Plank constant $\hbar_0$.

In order to investigate the evolution of the system, first, we
have to find stationary eigenstates of the unperturbed ($f_0=0$)
system, corresponding to the vicinity of the coupling resonance
$\omega_x=\omega_y$. At the second stage, we will use the Floquet
formalism when considering the time-periodic perturbation for $f_0
\neq 0$. Specifically, we construct the evolution operator in one
period $T$ of the perturbation, that allows one to study the
dynamics over many periods.

\subsection{Stationary states of the coupling resonance}

It is naturally to represent stationary states of the unperturbed
Hamiltonian $\hat H_s$,
\begin{equation}
\label{Hs}
\hat H_s=\hat H^0_x+\hat H^0_y-\mu xy
\end{equation}
in terms of the eigenstates of uncoupled ($\mu=0$) nonlinear
oscillators,
\begin{equation}
\label{psi_x_y}
\psi(x,\,y)=\sum_{n,m} c_{n,m}\psi^0_n(x)\psi^0_m(y).
\end{equation}
Here $\psi^0_n(x)$, $\psi^0_m(y)$ are the eigenfunctions of $\hat
H^0_x$, $\hat H^0_y$ (which will be calculated numerically), and
the coefficients $c_{n,m}$ satisfy to the following stationary
Schr\"odinger equation,
\begin{equation}
\label{stat eq}
Ec_{n,m}=(E_{n}+E_{m})c_{n,m}-
\mu\sum_{n',m'} x_{n,n'}y_{m,m'}c_{n',m'},
\end{equation}
with $E_{n}$ and $E_{m}$ as eigenvalues of the Hamiltonians $\hat
H^0_x$ and $\hat H^0_y$, respectively.

At the center of the coupling resonance
$\omega_{n_0}=\omega_{m_0}$ we have $n_0=m_0$ and
$\hbar_0\omega_{n_0}=E_{n_0}'$, $\hbar_0\omega_{m_0}=E_{m_0}'$.
Near to this resonance it is convenient to expand $E_n$ and $E_m$
in Tailor series up to second order terms. This allows one to
introduce new indexes $p=k+l$ and $k$ via the relations $n-n_0=k$
and $m-m_0=l$. Then our system (\ref{stat eq}) can be written in
the following form,
\begin{equation}
\label{system}
\matrix{
Ec_{k,p}=\left[\hbar_0\omega p+E_{n_0}''\left(k^2-pk+{p^2\over 2}
\right)\right]c_{k,p}-\cr
-\mu\Bigl(\dots + \sum_{k'} x_{k,k'}y_{p-k,-1-k'}\,c_{k',-1}+
\Bigr.\cr
\quad\;\;\,+ \sum_{k'} x_{k,k'}y_{p-k,-k'}\,c_{k',0}+ \cr
\Bigl. \qquad\qquad\;\:\,    + \sum_{k'} x_{k,k'}y_{p-k,1-k'}\,c_{k',1}+
\dots \Bigr)
}
\end{equation}
with $\omega \equiv \omega_{n_0}$. One should note that matrix
elements $x_{m,n}$ and $y_{m,n}$ of the coordinates $x$ and $y$
are equal to zero for transitions between the states of equal
parity. Therefore, the exact solution of the system (\ref{system})
is characterized by two independent sets of odd and even parity
eigenstates (for odd and even $p$ respectively).

\begin{figure}[tb]
\begin{center}
\mbox{\psfig{file=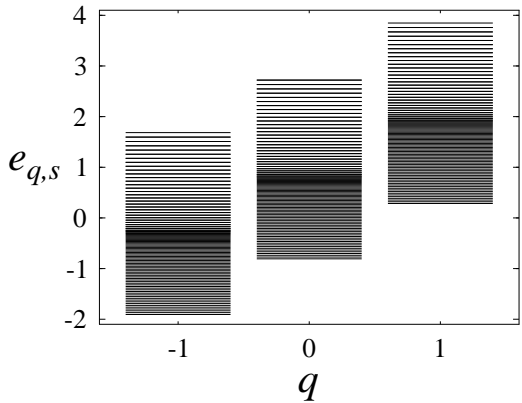,width=8cm,height=6cm}}
\end{center}
\caption{Energy spectrum of the system (\ref{system}) in
normalized units $e_{q,s}=E_{q,s}/\hbar_0\omega$ for
$\mu=10^{-4}$, $\hbar_0=1.77321\cdot 10^{-5}$ and $n_0=446$. Three
groups with 121 states in each group are shown.}
\label{spectrum}
\end{figure}

Below we consider the case of a small nonlinearity,
\begin{equation}
\label{condition}
\hbar_0\omega p\gg E_{n_0}''\left(k^2-kp+{p^2\over 2}\right).
\end{equation}
This allows us to characterize all states by the group number $q$
and by the index $s$ that stands for energy levels inside each
group. In fact, $q$ and $s$ are similar to {\it fast} and {\it
slow} classical variables characterizing the motion inside the
coupling resonance. Therefore, the energy in each group can be
written as $E_{q,s}=\hbar_0
\omega q+E^M_{q,s}$, where $E^M_{q,s}$ is the Mathieu-like
spectrum of one group.

Numerical data for a fragment of the energy spectrum are shown in
Fig.\ref{spectrum}. One can see that the spectrum consists of
series of energy levels, shifted one from another by the value
$\hbar_0\omega$. Note that the structure of the energy spectrum in
each group is typical for a quantum nonlinear resonance
\cite{BVI87}. Lowest levels are practically equidistant with the
spacing equal to $\hbar_0\tilde
\omega$, where $\tilde \omega$ is the classical frequency of small
phase oscillations at the coupling resonance. Accumulation points
correspond to classical separatricies, and all energy levels
inside separatricies are non-degenerate. The states slightly above
separatricies are quasi-degenerate due to the symmetry of a
rotation in opposite directions.

\begin{figure}[tb]
\begin{center}
\vspace{-1cm}
\mbox{\psfig{file=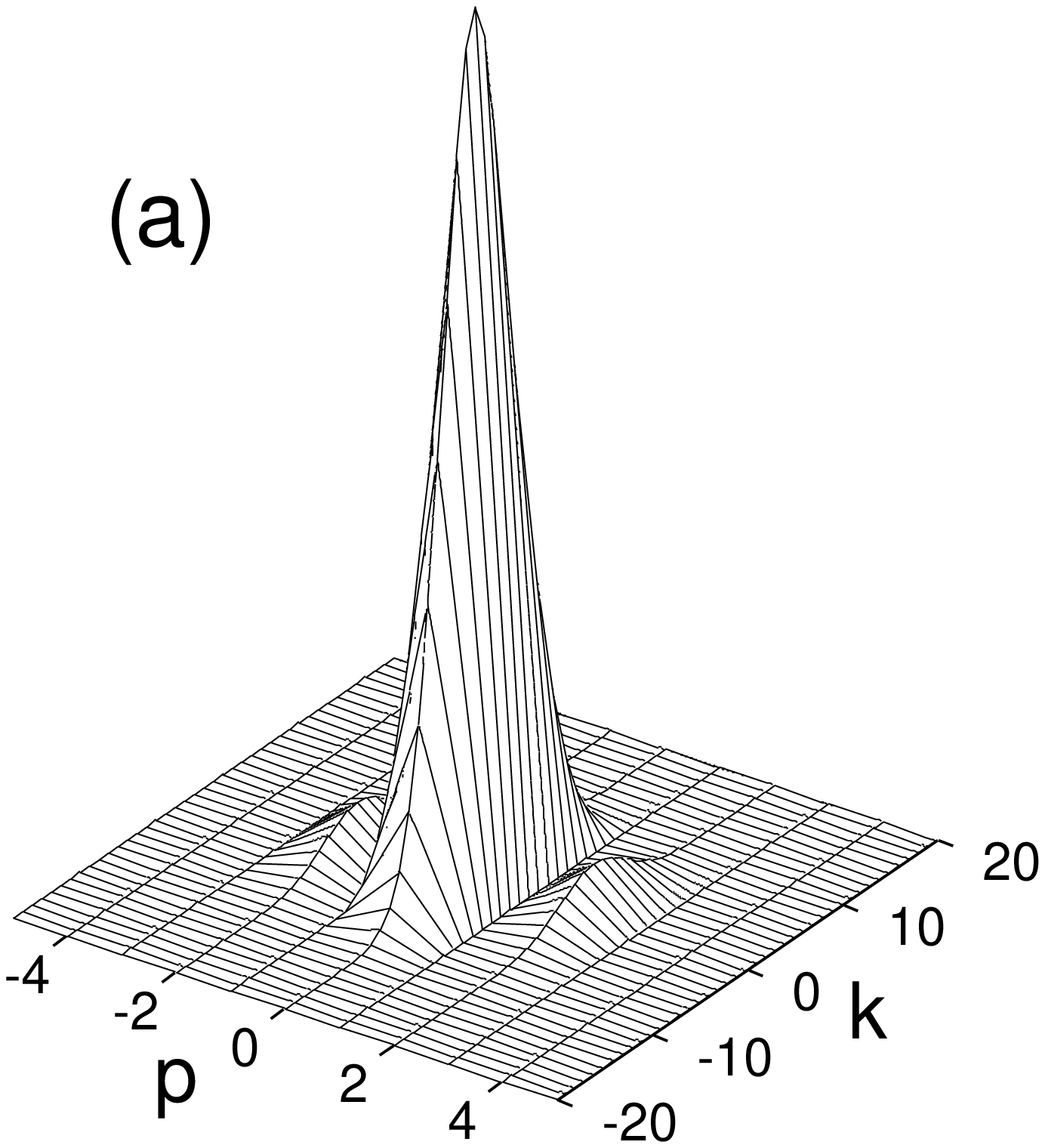,width=4.3cm,height=4.cm}
\psfig{file=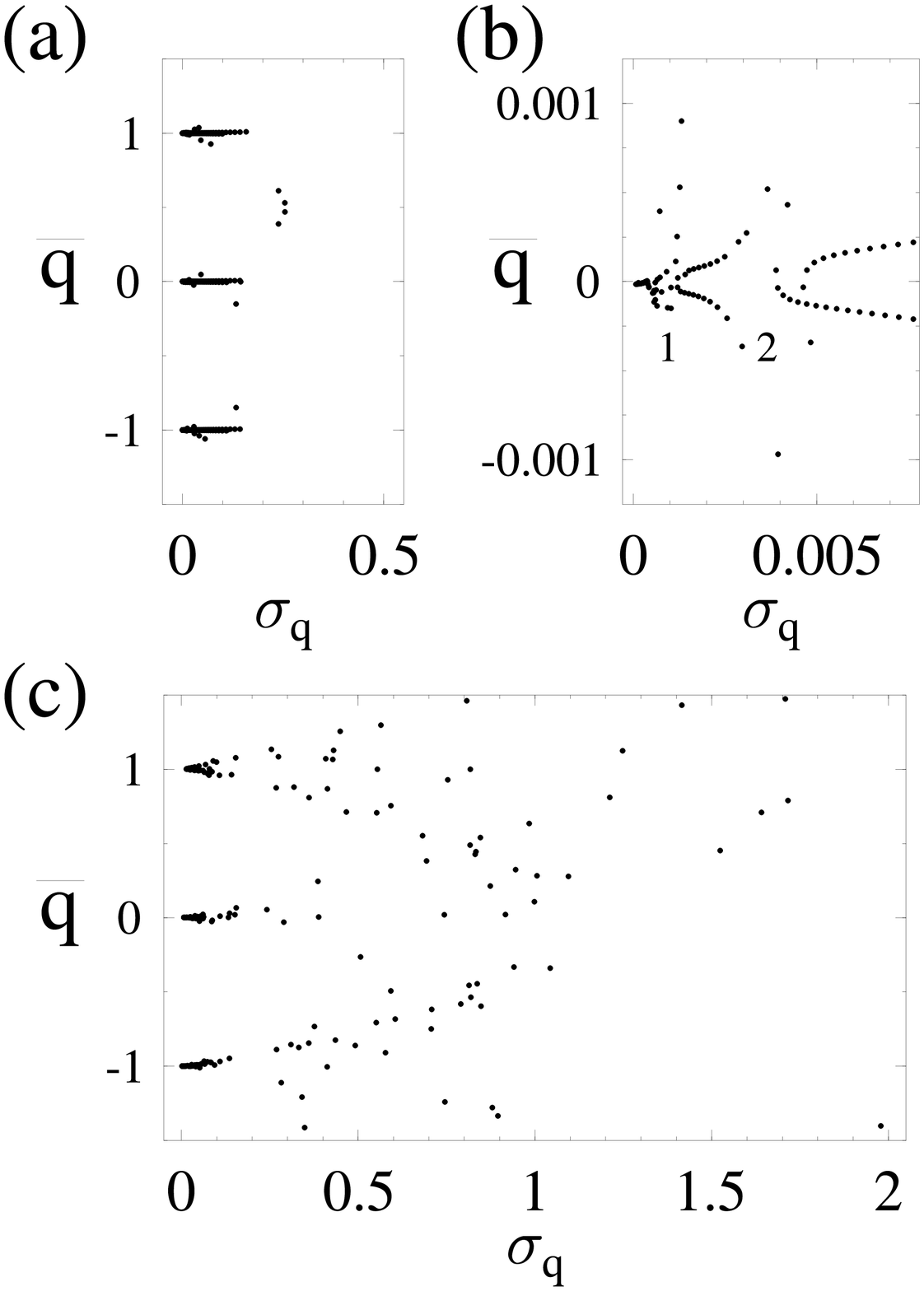,width=4.3cm,height=4.cm}}

\mbox{\psfig{file=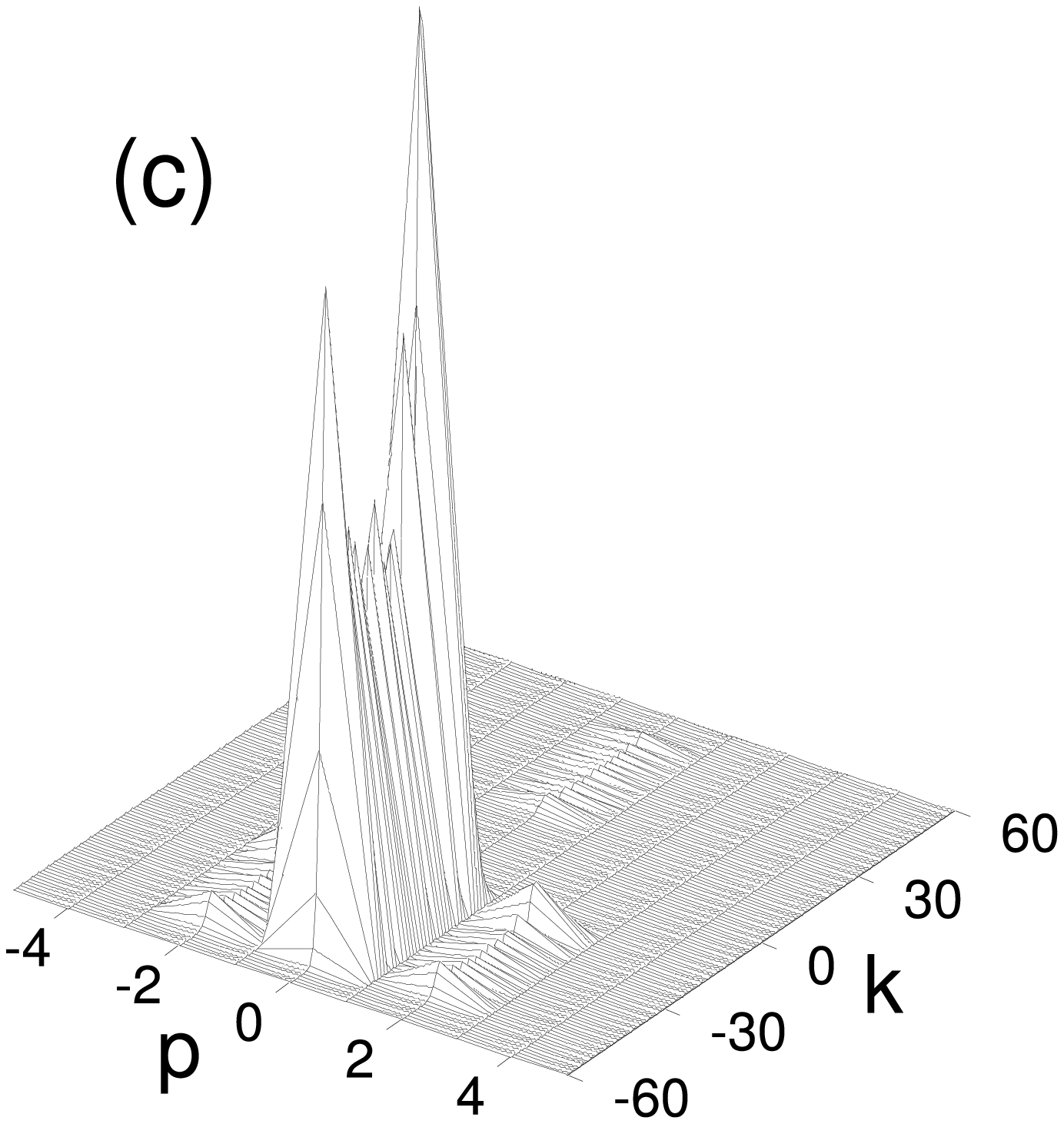,width=4.3cm,height=4.cm}
\psfig{file=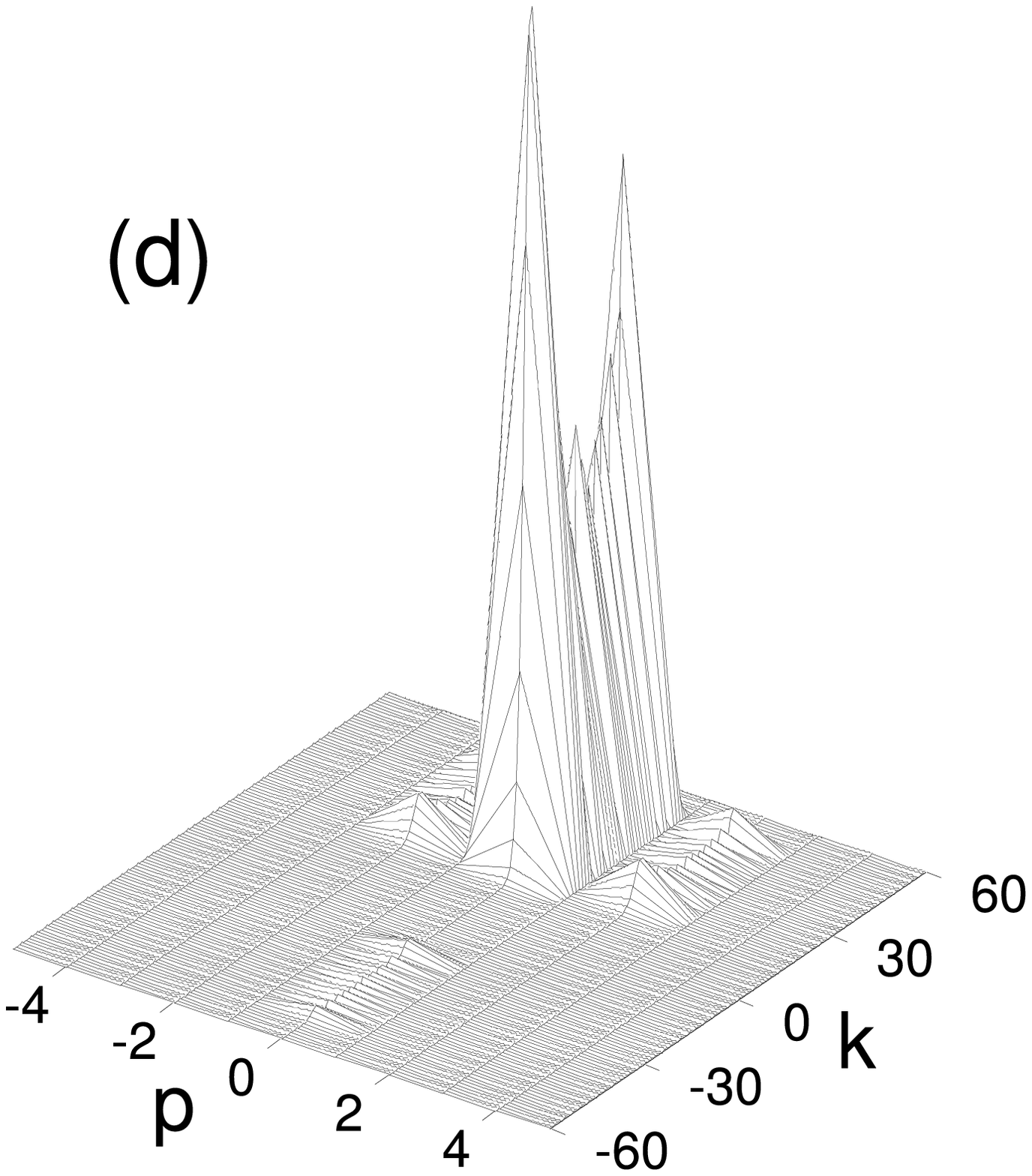,width=4.3cm,height=4.cm}}
\end{center}
\caption{Probability $|c_{k,p}|^2$ for eigenstates at the coupling
resonance for $q=0$, $\mu=10^{-4}$ and different values of $s$. a)
the lowest level (ground state) for $s=0$ ; b) near-separatrix
level, $s=30$; c), d) above-separatrix levels, $s=\pm 35$.}
\label{func}
\end{figure}

Typical structure of eigenstates for different $s$ is shown in
Fig.\ref{func}. Note that ground states in each group correspond
to $s=0$, and next stationary states, reordered according to the
energy increase, are labelled by $1,-1,2,-2,\dots$ etc. The
eigenstates inside the resonance are symmetrical with respect to
$k=0$. One can see that the main maximum corresponds to $p=0$,
although there are small additional maximums at $p=\pm 2,\pm
4,\dots$. The degree of delocalization in the $k$-space for
eigenstates inside the coupling resonance strongly depends on the
energy of eigenstates. Specifically, the closer the energy to that
corresponding to the separatrix, the more delocalized is the
eigenstate. Above the separatrix all eigenstates are characterized
by the maximums of probability located symmetrically with respect
to $k=0$ for states $s$ and $-s$, see Fig.\ref{func}(c),(d).

\subsection{Evolution matrix}

Now we consider the dynamics of our model in the presence of the
external two-frequence perturbation acting on the $x$-oscillator.
The frequencies $\Omega_1$ and $\Omega_2$ are commensurable, so
that the external force is periodic with the period $T=i T_1=j
T_2$, where $T_1=2\pi/\Omega_1$, $T_2=2\pi/\Omega_2$ and $i$, $j$
are integer. The initial conditions were taken for the system to
be about half-way between the two driving resonances, $\omega =
(\Omega_1+\Omega_2)/2$.

Since the Hamiltonian (\ref{ham}) is periodic in time, in
accordance with the Floquet theory the solution of the
non-stationary Schr\"odinger equation can be written in the
following form,
\begin{equation}
\label{Flok}
\psi(x,y,t)=\exp\left( -{i\varepsilon_Qt\over \hbar_0}\right)
u_Q(x,y,t).
\end{equation}
Here $u_Q(x,y,t)=u_Q(x,y,t+T)$ is the {\it quasienergy function}
with the corresponding {\it quasienergy} (QE) $\varepsilon_Q$. The
QE functions and quasienergies are, in fact, the eigenfunctions
and eigenvalues of the evolution operator $\hat U(T)$ that
describes the evolution of the system within one period of the
external perturbation,
\begin{equation}
\label{ev_1}
\hat U(T)u_Q(x,y)=\exp\left( -{i\varepsilon_QT\over \hbar_0}\right)
u_Q(x,y).
\end{equation}
Since we are interested in wave functions only in discrete times
$N\cdot T$ with $N$ integer, we omitted the argument $t$ in
Eq.(\ref{ev_1}).

In order to construct the evolution operator, we represent the QE
functions as follows,
\begin{equation}
\label{ev_2}
u_Q(x,y)=\sum_{q,s} A^Q_{q,s}\psi_{q,s}(x,y).
\end{equation}
Here the functions $\psi_{q,s}(x,y)$ are eigenstates of the
stationary Hamiltonian $\hat H_s$ (see Eq.(\ref{Hs})), and the
coefficients $A^Q_{q,s}$ are the eigenvectors of the operator
$\hat U$ in the representation of $\hat H_s$. These eigenvectors
can be found by a direct diagonalization of the corresponding
matrix $U_{q,s;q',s'}$.

To obtain the matrix $U_{q,s;q',s'}$ we have used the following
procedure. Let the evolution operator $\hat U$ act on an initial
state $C^{\left(q_0,s_0\right)}_{q,s}(0)
=\delta_{q,q_0}\delta_{s,s_0}$. Then the wave function
$C^{(q_0,s_0)}_{q,s}(T)$ at time $T$ forms the column of the
evolution operator matrix,
\begin{equation}
\label{ev_3}
\matrix{
U_{q,s;q',s'}(T)C^{(q_0,s_0)}_{q',s'}(0)=U_{q,s;q_0,s_0}(T)
%=\cr
=C^{(q_0,s_0)}_{q,s}(T).}
\end{equation}
Repetition of this procedure for different initial states
$C^{(q',s')}_{q,s}(0)=\delta_{q,q'}\delta_{s,s'}$ allows one to
find the whole matrix $U_{q,s;q',s'}(T)$. As a result, the wave
function $C^{(q_0,s_0)}_{q,s}(T)$ can be computed numerically by
integration of the nonstationary Schr\"odinger equation in the
presence of the time-dependent perturbation,
\begin{equation}
\label{time eq}
\matrix{
i\hbar_0 \dot C_{q,s}=\left( \hbar_0 \omega q+E^M_{q,s}\right)
C_{q,s}- \cr -f_0\sum_{q',s'} x_{q,s;q',s'}\left(
\cos\Omega_1t+\cos\Omega_2t \right)C_{q',s'}}
\end{equation}
By introducing the slow amplitude $b_{q,s}(t)$ via the
transformation
\begin{equation}
\label{change}
C_{q,s}(t)=b_{q,s}(t)\exp\left[ -i\left( q \omega
+E^M_{q,s}/\hbar_0 \right) t\right]
\end{equation}
one can obtain,
\begin{equation}
\label{reson}
\matrix{
i\hbar_0 \dot b_{q,s}=-f_0\cos\left({\delta \Omega \over
2}t\right) \times \cr
\times \sum_{s'}
\left[ x_{q,s;q+1,s'}\,b_{q+1,s'}\,e^{-i\bigl(E^M_{q+1,s'}-E^M_{q,s}
\bigr)t/\hbar_0 }+ \right. \cr
\left. \qquad\; +\,x_{q,s;q-1,s'}\,b_{q-1,s'}\,e^{-i\bigl
( E^M_{q-1,s'}-E^M_{q,s}\bigr)t/\hbar_0 } \right],}
\end{equation}
where $\delta \Omega = \Omega_1-\Omega_2$. In the resonance
approximation we keep in Eq.(\ref{reson}) only the most important
slowly oscillating terms with $q'=q\pm 1$.

%\begin{figure}[tb]
%\begin{center}
%\mbox{\psfig{file=m_el_new.ps,width=8cm,height=8cm}}
%\end{center}
%\caption{The matrix elements $|x_{q,s;q\pm 1,s'}|$, which define the
%transition probability through the coupling resonance (for the
%same parameters as at Fig.\ref{spectrum}).}
%%  $\hbar_0=1.77321\cdot 10^{-5}$ and $n_0=446$.}
%\label{matr_el}
%\end{figure}

\begin{figure}[tb]
\begin{center}
\mbox{\psfig{file=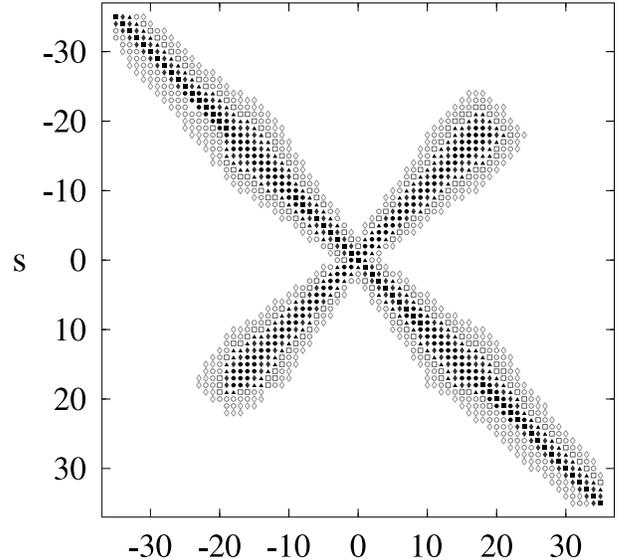,width=8cm,height=8cm}}
\end{center}
\vspace{-1.0cm}
\caption{Matrix elements $|x_{q,s;q\pm 1,s'}|$ that define the
transition probability along the coupling resonance (for the same
parameters as in Fig.\ref{spectrum}). Different values of the
matrix elements are shown by grayscaled symbols. Black squares at
the diagonal of the matrix have the values six orders of magnitude
higher than those of matrix elements labeled by grey rhombuses.}
\label{matr_el}
\end{figure}

The matrix elements $x_{q,s;q\pm 1,s'}$ in Eq.(\ref{reson})
correspond to transitions between the states $s$ and $s'$ from
neighbor groups $q$ and $q\pm 1$. Figure \ref{matr_el} illustrates
relative amplitudes of the matrix elements $|x_{0,s;1,s'}|$. In
accordance with our numeration of the states, matrix elements at
the center of Fig.\ref{matr_el} correspond to transitions between
the lowest states in each group, and matrix elements at the
corners define the transitions between the states above
accumulation points. The latter elements quickly decrease with an
increase of the difference $|s-s'|$.

The \lq\lq cross\rq\rq{} at the center of Fig.\ref{matr_el} where
matrix elements are relatively large, corresponds to a transition
between separatrix states. The important point is that the
transition between such states of neighbor groups (along the
coupling resonance) is much stronger than those between other
states. This phenomenon is analogous to the quantum diffusion
inside a separatrix, which was observed in \cite{demi} for a
degenerate Hamiltonian system.

As a result, numerical procedure for computing the dynamics of our
model is as follows. First, we solve Eqs.(\ref{reson}) and
construct the evolution matrix $U_{q,s;q',s'}(T)$ by making use of
Eq.(\ref{change}). Then, direct diagonalization of this matrix
yields the eigenvalues $\varepsilon_Q$ and the eigenvectors
$A^Q_{q,s}$. Once the eigenvalues $\varepsilon_Q$ and eigenvectors
$A^Q_{q,s}$ are obtained, one gets the evolution operator for one
period,
\begin{equation}
\label{ev_4}
U_{q,s;q',s'}(T)=\sum_{Q} A^Q_{q,s} {A^{Q^{\,*}}_{q',s'}} exp\left
( -{i\varepsilon_QT\over \hbar_0}\right).
\end{equation}
By raising $U_{q,s;q',s'}(T)$ to the $N$-th power and using the
ortogonality condition for the eigenvectors $A^Q_{q,s}$, one can
obtain the evolution operator that propagates the system over $N$
periods of the external perturbation,
\begin{equation}
\label{ev_5}
U_{q,s;q',s'}(NT)=\sum_{Q} A^Q_{q,s} {A^{Q^{\,*}}_{q',s'}}
exp\left( -{i\varepsilon_QNT\over
\hbar_0}\right).
\end{equation}

%%%%%%%%%%%%%%%%%%%%%%%%%%%%%%%%%%%
\subsection{Numerical data}

As was shown above, in our model the Arnol'd diffusion occurs
along the coupling resonance, or, the same, in $q$-space. This
means that a wave packet initially localized at $q=0$, spreads
diffusively in time. In order to observe this dynamics, below we
introduce specific variables that characterize global structure of
wave packets. But, first, we discuss the structure of QE
eigenstates (in terms of these new variables) since it helps us to
understand the mechanism of quantum Arnol'd diffusion.

\begin{figure}[tb]
\begin{center}
\mbox{\psfig{file=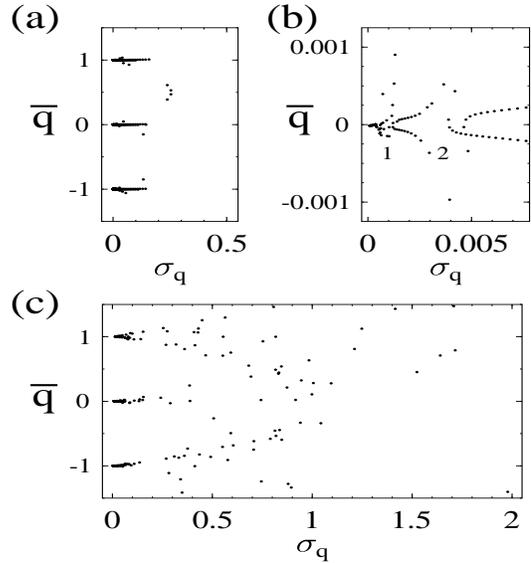,width=8cm,height=8cm}}
\end{center}
\caption{Relation between $\bar q$ and $\sigma_q$ for the
QE functions $A^Q_{q,s}$ in the region $q=0,\pm 1$ for different
coupling constants: a,b) $\mu=3\cdot 10^{-5}$, c) $\mu=10^{-4}$.
Each point corresponds to a specific QE function.
Fig.\ref{cluvs}(b) shows the scaled-up fragment of the central
part of Fig.\ref{cluvs}(a).}
\label{cluvs}
\end{figure}

In the $q$-space each of QE functions $A^Q_{q,s}$ can be globally
characterized by the ``mean position" $\bar q$ and by the variance
$\sigma_q^2$ determined as follows \cite{BVI87,toda},
\begin{equation}
\label{sigma}
\bar q=\sum_{q}q\sum_{s}|A^Q_{q,s}|^2, \,\,\,\,\,\,\,\,\,
\sigma_q^2=\sum_{q}(q-\bar q)^2\sum_{s}|A^Q_{q,s}|^2.
\end{equation}
Then, it is convenient to plot $\bar q$ versus $\sigma_q$ for all
eigenfunctions, see Fig.\ref{cluvs}. For small values of the
coupling $\mu$, see Fig.\ref{cluvs}(a), the QE functions have a
very small variance that means a strong localization in the
$q$-space.

More details are seen in  Fig.\ref{cluvs}(b) which shows a
magnified fragment of Fig.\ref{cluvs}(a). One can see different
groups of QE functions, characterizing specific relations between
$\bar q$ and $\sigma_q$. First group consists of those eigenstates
whose energies are close to the ground state (with a very small
variance, $\sigma_q<10^{-3}$). Another group is characterized by
an irregular dependence of $\bar q$ on $\sigma_q$ (region (1) in
Fig.\ref{cluvs}(b)). The states which belong to this group are
chaotic separtrix eigenstates. Regular dependences $\bar
q(\sigma_q)$ for $\sigma_q>10^{-3}$ correspond to under-separatrix
states with $s>0$ and $s<0$. For very large $s$ this regular
structure is destroyed due to the influence of two driving
resonances (not shown in Fig.\ref{cluvs}(b)). Irregular spread of
points in the region (2) reflects chaos in the inner region of the
coupling resonance, that arises due to two secondary resonances
$\hbar_0 \delta\Omega /2 = 3E''_{n_0}(s+1/2)$. With an increase of
$\mu$ (see Fig.\ref{cluvs}(c)), the regular structure of QE
functions disappears. This means that many of eigenstates are
chaotic. However, the variance $\sigma_q^2$ remains limited, thus,
demonstrating that in the $q$ direction the eigenstates are {\it
localized}. The fact that many points in Fig.\ref{cluvs}(c) are
distributed incidentally, should be treated as the manifestation
of quantum chaos.

%%%%%%%%%%%%%%%%%%%%%%%%%%%%%%%%%%%%%%%%%%%%%%%

Now we discuss numerical results for the dynamics of our model.
The evolution of any initial state $C_{q',s'}(0)$ can be computed
using the evolution matrix $U_{q,s;q',s'}(NT)$,
\begin{equation}
\label{ev_6}
C_{q,s}(NT)=\sum_{q',s'}U_{q,s;q',s'}(NT)C_{q',s'}(0).
\end{equation}
We repeat again that our numerical data refer to the regime when
the values of $\mu$ and $f_0$ are small enough, so that main three
resonances are not overlapped.

\begin{figure}[tb]
\begin{center}
\mbox{\psfig{file=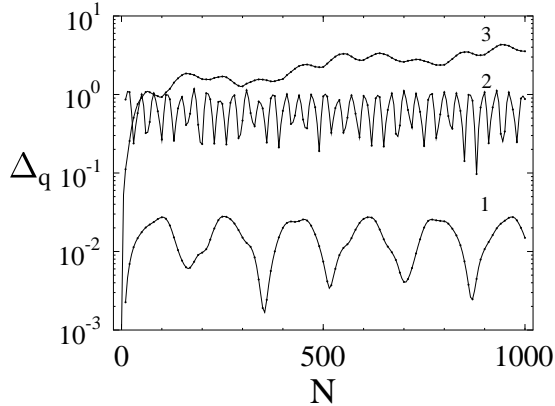,width=8cm,height=6cm}}
\end{center}
\caption{Normalized variance $\Delta_q$ of the energy
versus the rescaled time $N$ for 3 different initial states (for
$\mu=1.25\cdot 10^{-4}$ and $f_0=1.25\cdot 10^{-6}$). Curves 1-3
correspond to an initial state near the center of the coupling
resonance, above the separatrix, and from the separatrix layer,
respectively.}
\label{comp}
\end{figure}

Quantum dynamics for different initial conditions is shown in
Fig.\ref{comp}. Here we show typical dependencies of the variance
of the energy $\overline {(\Delta H)^2}=\hbar_0^2\omega^2\Delta_q$
in normalized units, versus time $t$ measured in the number $N$ of
periods of the external perturbation. The quantity $\Delta_q$ is
defined similar to that for the QE eigenstates,
$\Delta_q=\sum_{q}(q-\bar q)^2\sum_{s}|C_{q,s}|^2$ where $\bar
q=\sum_{q}q\sum_{s}|C_{q,s}|^2$.

The data clearly show a different character of the evolution for
three initial states taken from below and above the separatrix, as
well as from the separatrix layer. For the states taken from the
center of the resonance and well above the separatrix, the
variance $\overline {(\Delta H)^2}$ quasi-periodically oscillates,
in contrast with the separatrix state. In the latter case, after a
short stochastization time the variance of the energy increases
linearly in time, thus manifesting a diffusion-like spread of the
wave packet.

\begin{figure}[tb]
\begin{center}
\mbox{\psfig{file=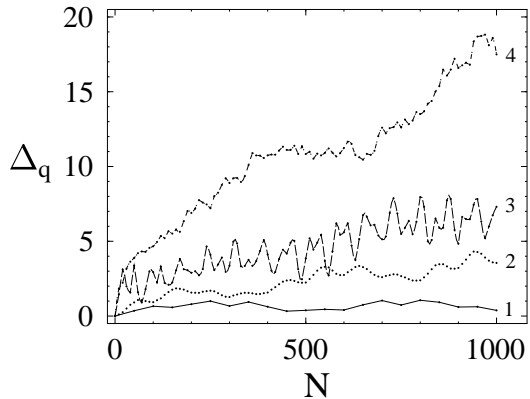,width=8cm,height=6cm}}
\end{center}
\caption{Time dependence of the variance of the energy
for separatrix initial states and different values of coupling
parameter $\mu$: 1) $\mu=10^{-4}$, 2) $\mu=1.25 \cdot 10^{-4}$, 3)
$\mu=1.75
\cdot 10^{-4}$, 4) $\mu=2.25 \cdot 10^{-4}$.}
\label{dyn}
\end{figure}

More results are presented in Fig.\ref{dyn} where a diffusion-like
increase of the energy is shown for separatrix initial states and
different values of $\mu$. One can see that linear increase of the
variance $\Delta_q(N)$ is typical, and it occurs after a short
time which is associated with the time of a fast spread of packet
over the separatrix layer in transverse direction. Such a
behaviour is typical for the classical Arnol'd diffusion. The data
allows one to determine the diffusion coefficient as
$D={\overline{(\Delta H)^2}(N)}/(N\cdot T)$, by making use the fit
to a linear dependence $\Delta_q(N)$ for $N > 50$.

\begin{figure}[tb]
\begin{center}
\mbox{\psfig{file=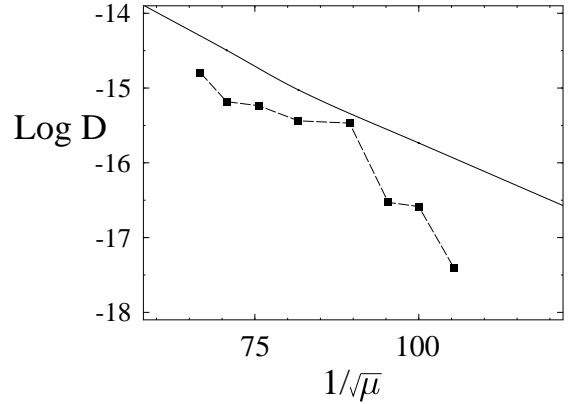,width=8cm,height=6cm}}
\end{center}
\vspace{-0.3cm}
\caption{Quantum (squares) and classical (solid line) diffusion
coefficients for different values of $\mu$.}
\label{coeff}
\end{figure}

We have calculated separately quantum and classical diffusion
coefficients and found that the quantum Arnol'd diffusion roughly
corresponds to the classical one. However, the data clearly
indicate that the quantum diffusion is systematically weaker than
the classical Arnol'd diffusion, see Fig.\ref{coeff}.

One should stress that the quantum Arnol'd diffusion takes places
only in the case when the number $M_s$ of energy stationary states
in the separatrix layer is relatively large. For the first time,
this point was noted by Shuryak \cite{S76} who studied the
quantum-classical correspondence for nonlinear resonances. In this
connection we have to estimate the number of the energy states
that occupy the separatrix layer.

However, it is a problem to make reliable analytical estimates of
the width of the separatrix layer for our model. As was shown in
\cite{Vech}, in the case of many-frequency perturbation,
secondary resonances play the dominant role in the formation of
chaos in a separatrix layer. Indeed, as was shown above, the role
of such resonances is quite strong in our case of the
two-frequency perturbation. For this reason, we have performed a
direct numerical calculation of the width of stochastic separatrix
layer for the classical Hamiltonian, and used these results in
order to estimate the number of separatrix energy levels in the
corresponding energy interval.

We have found that for $\mu>1.25\cdot 10^{-4}$ the number $M_s$ of
stationary states in the separatrix chaotic layer is more than 10,
therefore, one can speak about a kind of stochastization in this
region. On the other hand, with a decrease of the coupling (see
data in Fig.\ref{coeff} for $1/\sqrt{\mu}>100$), the number $M_s$
decreases and for $\mu \approx 3\cdot 10^{-5}$ it is of the order
one. For this reason the last right point in Fig.\ref{coeff}
corresponds to the situation when classical chaotic motion along
the coupling resonance is completely suppressed by quantum effects
(the so-called ``Shuryak border" \cite{S76}).

\subsection{Dynamical localization}

Since the diffusive motion along the coupling resonance is
effectively one-dimensional, one can naturally expect the
Anderson-like localization. We have already noted that the
variance of QE eigenstates of the evolution operator is finite in
the $q$-space. This means that eigenstates are localized, and the
wave packet dynamics in this direction has to reveal the
saturation of the diffusion. More specifically, we expect that the
linear increase of the variance of the energy ceases, after some
characteristic time.

This effect known as the {\it dynamical localization}, has been
discovered in \cite{CCIF79,CIS81} for the kicked rotor, and was
studied later in different physical models (see, for example,
\cite{LL92} and references therein). One should note that the
dynamical localization is, in principle, different from the
Anderson localization, since the latter occurs for models with
random potentials. In contrast, the dynamical localization happens
in dynamical (without any randomness) systems, and is due to
interplay between (week) classical diffusion and (strong) quantum
effects.

\begin{figure}[tb]
\begin{center}
\mbox{\psfig{file=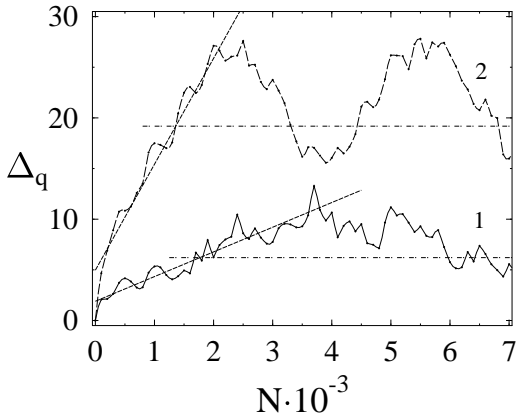,width=8cm,height=6cm}}
\end{center}
\caption{Dynamical localization of wave packets inside the
separatrix layer of the coupling resonance.  The normalized
variance of the energy is shown for different values of the
coupling parameter: (1) $\mu=1.5\cdot 10^{-4}$, (2) $\mu=2.25\cdot
10^{-4}$. Horizontal dashed lines indicate the mean value
$\overline{\Delta_q}$.}
\label{limit}
\end{figure}

In order to observe the dynamical localization in our model (along
the coupling resonance inside the separatrix layer), one needs to
study long-time dynamics of wave packets. Our numerical study for
large times $N \approx 10^4$ have revealed that after some time
$t\sim t_0 \approx 10^3\cdot T$, the diffusionlike evolution stops
for all range of coupling parameter $\mu$. Instead, for larger
times, the variance $\Delta_q$ starts to oscillate around the mean
value $\overline{\Delta_q}$. Period of such oscillations depends
on $\mu$ non-monotonically and vary from $10^{3}\cdot T$ to
$10^{4}\cdot T$. Two examples of such a long-time dynamics are
given in Fig.\ref{limit}.

\begin{figure}[tb]
\begin{center}
\mbox{\psfig{file=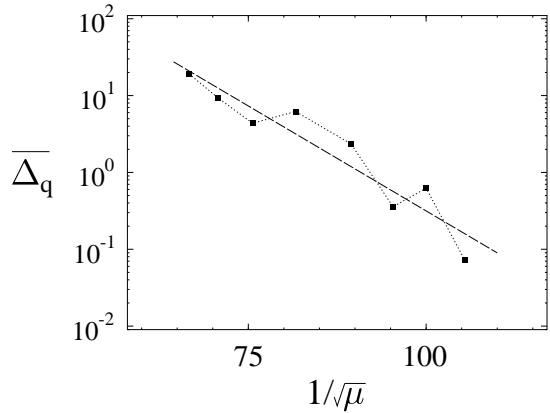,width=8cm,height=6cm}}
\end{center}
\caption{The dependence of $\overline{\Delta_q}$
on $1/\sqrt{\mu}$.}
\label{localiz}
\end{figure}

One can argue that the localization length is of the order of the
width of wave packet after the saturation of the diffusion
\cite{CIS81,CIS81}. Therefore, the
localization length $l_s$ can be associated with the square root
of $\overline{\Delta_q}$.  We have numerically found the
exponential dependence $\overline{\Delta_q}$ on $1/\sqrt{\mu}$,
see Fig.\ref{localiz}. This result is not surprising because
$\log{D}$ is proportional to $1/\sqrt{\mu}$ (see Fig.\ref{coeff})
and $t_0$ is practically independent on $\mu$.

\begin{figure}[tb]
\begin{center}
\mbox{\psfig{file=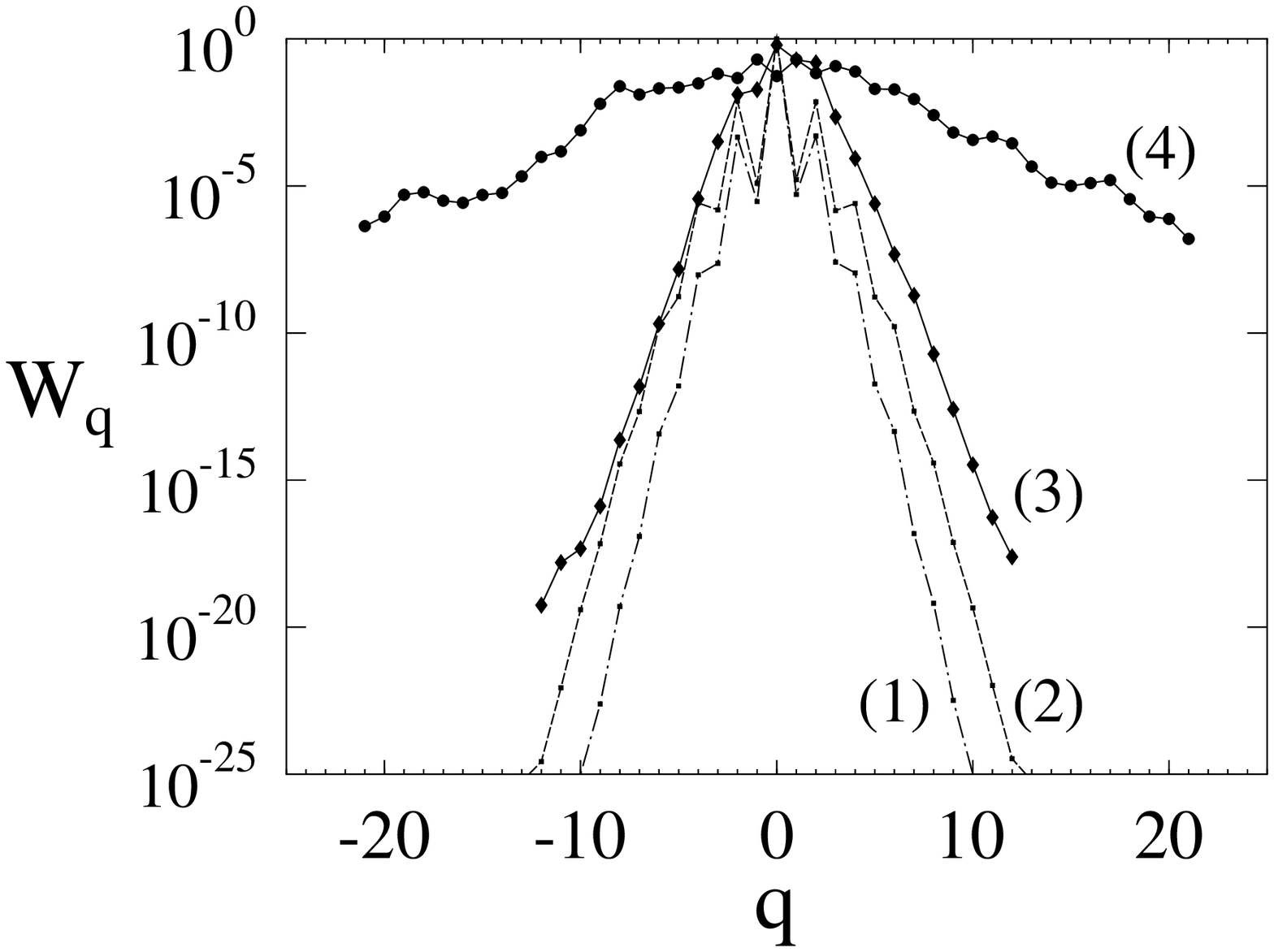,width=8cm,height=6cm}}
\end{center}
\caption{Distribution of the probability $W_q$, averaged over
time $t=10^4\cdot T$, versus the group number $q$: (1) $\mu=0$,
$f_0=10^{-6}$; (2) $\mu=0$, $f_0=2\cdot 10^{-6}$; (3)
$\mu=10^{-4}$, $f_0=10^{-6}$; (4) $\mu=2\cdot 10^{-4}$,
$f_0=2\cdot 10^{-6}$.}
\label{decreas}
\end{figure}

It should be noted that the spread of packets in $k$ and $l$ space
occurs even without the coupling between two resonances, due to
the influence of the time-dependent perturbation (see Section
III-A). For this reason one should compare this spread to that
determined by the Arnol'd diffusion. Our additional study clearly
show that these two effects are very different, see
Fig.\ref{decreas}.

Here, we plotted the profile $W_q=\sum_s |C_{q,s}|^2$ of the wave
packet after the saturation, versus the group number $q$. This
figure demonstrates the main effect of the Arnol'd diffusion along
the coupling resonance. One can see that in all cases there is an
exponential localization of packets in the $q$-space. This allows
to introduce the localization length defined from the decrease of
the probability in the tails of packets. The data illustrate a
strong increase of the localization length in the presence of the
coupling between two oscillators, in comparison with the case of
completely independent oscillators for $\mu=0$ ($q=k+l$ and
$l=0$).

\section{Summary}

We have studied the Hamiltonian system of two coupled nonlinear
oscillators, one of which is under the influence of the
time-dependent perturbation with two commensurate frequencies. In
the classical description, the separatrix of the coupling
resonance is destroyed due to the perturbation, and the Arnol'd
diffusion occurs along this resonance inside a narrow stochastic
layer. Our numerical data performed for the quantum analog of the
system, allow us to make the following conclusions about the
properties of the quantum Arnol'd diffusion.

By studying the quasienergy eigenstates of the evolution operator,
we have found an irregular structure of those eigenstates which
correspond to the stochastic layer. These eigenstates turned out
to be exponentially localized along the coupling resonance, with
the localization length, strongly enhanced in comparison with the
case of non-coupled oscillators.

The study of wave packet dynamics for different initial states
have revealed a diffusion-like spread of packets along the
coupling resonance, if initial states correspond to the stochastic
layer. We have found that the dependence of the diffusion
coefficient on model parameters, roughly follows the classical
dependence. However, the quantum diffusion is systematically
slower than the classical one. This fact is obviously due to the
influence of quantum effects.

It should be stressed that the quantum Arnol'd diffusion occurs in
a deep semiclassical region, specifically, for the case when the
number $M_s$ of chaotic eigenstates inside the stohastic layer is
sufficiently large (of the order of 10 or larger). With a decrease
of the coupling parameter, the diffusion coefficient strongly
decreases, and for $M_s \leq 1$ the diffusion disappears.
Therefore, we can see how strong quantum effects destroy the
diffusive dynamics of wave packets.

Another manifestation of quantum effects is the dynamical
localization that persists even for large $M_s$. Specifically, we
have observed that the quantum diffusion occurs only for finite
(although large) times. On a larger time scale the diffusion
ceases, and after some characteristic time it terminates. This
effect is similar to that discovered in the kicked rotor model
\cite{CCIF79} and studied later in other physical systems (see,
for example, \cite{LL92} and references therein). However, in our
case the dynamical localization arises for a weak chaos inside the
separatrix layer, in contrast to previous models with a strong
(global) chaos in the classical description.

Our results may find a confirmation in experiments on one-electron
dynamics in 2D semiconductor quantum billiards, where the charged
particle motion is determined by the Hamiltonian of the type
(\ref{ham}), (\ref{hatH^0_x}). It is also possible that the
quantum Arnol'd diffusion occurs in nuclear dynamics of complex
molecules, driven by laser fields \cite{Perry}.

\section*{Acknowledgments}
The authors are thankful to B.~Chirikov for stimulating
discussions. We also thank D.~Kamenev for the help in performance
of some calculations on the early stage of our work. We wish to
thank D.~Leitner and P.~Wolynes for attracting our attention to
the reference \cite{LW97}. This work was supported by grants RFBR
No.~01-02-17102, and by the Ministry of Education of Russian
Federation No.~E00-3.1-413 and \lq\lq Universities of
Russia\rq\rq. FMI acknowledges the support by CONACyT (Mexico)
Grant No.~34668-E.

%*****************************************

\end{multicols}
\end{document}